\documentclass[twocolumn, pra, amsmath, amssymb, unsortedaddress,
superscriptaddress, showpacs]{revtex4}

\usepackage{latexsym, epsfig, graphicx}
\usepackage{dcolumn}
\usepackage{bm}
\usepackage{color} 
\usepackage{graphicx} \usepackage{subfigure} 
\newcommand{\be}{\begin{equation}} \newcommand{\ee}{\end{equation}}
\newcommand{\bea}{\begin{eqnarray}} \newcommand{\eea}{\end{eqnarray}}

\begin{document}

\title{Oscillatory pairing of fermions in spin-split traps}

\author{Kuei Sun} \affiliation{Department of Physics, University of
  Illinois at Urbana-Champaign, Urbana, Illinois 61801, USA} \author{Julia
  S. Meyer} \affiliation{SPSMS, UMR-E CEA/UJF-Grenoble 1, INAC,
  Grenoble, F-38054, France} \affiliation{Department of Physics, Ohio State University, Columbus, Ohio 43210, USA} \author{Daniel
  E. Sheehy} \affiliation{Department of Physics and Astronomy, Louisiana State
  University, Baton Rouge, Louisiana 70803, USA} \author{Smitha Vishveshwara}
\affiliation{Department of Physics, University of Illinois at
  Urbana-Champaign, Urbana, Illinois 61801, USA} \date{March 11, 2011}


\begin{abstract}
As a means of realizing oscillatory pairing between fermions, we
study superfluid pairing between two fermion \lq\lq spin\rq\rq\
species that are confined to adjustable spin-dependent trapping
potentials. Focusing on the one-dimensional limit, we find that
with increasing separation between the spin-dependent traps the
fermions exhibit distinct phases, including a fully paired phase,
a spin-imbalanced phase with oscillatory pairing, and an unpaired
fully spin-polarized phase. We obtain the phase diagram of
fermions in such a spin-split trap and discuss signatures of these
phases in cold-atom experiments.
\end{abstract}

\maketitle

\begin{figure}[t]
  \centering
  \includegraphics[width=7.5cm]{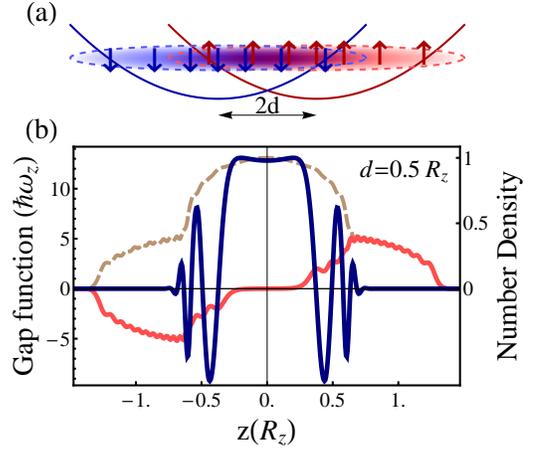}
  \caption{(Color online) (a) Illustration of our proposed spin-split
    trap setup showing separate trapping potentials for two fermion
    species, $\uparrow$ and $\downarrow$. (b) Numerically-determined
    spatial profile of the pairing gap $\Delta(z)$ (solid blue curve,
    axis on left-hand side of graph), total density $\rho(z)$, and spin
    imbalance $M(z)$ (dashed light-brown
    and solid red curves, respectively, axis on right-hand side
    of graph, normalized by $\max_z[\rho]$), showing oscillatory pairing along with a local
    imbalance.}
  \label{fig:f01}
\end{figure}

\section{Introduction}
 The idea that Cooper pairing in the presence of a density
imbalance of two interacting fermion species naturally yields
oscillatory pairing correlations in real space was put forth
decades ago. However, to date, this phenomenon, known as
Fulde-Ferrell-Larkin-Ovchinnikov (FFLO) pairing~\cite{FF,LO}, has
not been conclusively observed. (Related effects have been clearly
seen in superconductor-ferromagnet hybrid systems where the
proximity-induced pair correlations in the ferromagnet exhibit
oscillations~\cite{Buzdin05}.) In recent years, atomic physics
experiments have explored paired fermion superfluidity in cold
atomic gases~\cite{Bloch,Giorgini,Ketterle},  a new setting for
the observation of FFLO pairing correlations under a density
imbalance between the two \lq\lq spin\rq\rq\ species---a
possibility that has inspired a large amount of recent theoretical
and experimental activity~\cite{Radzihovsky}. Much of the
excitement follows from the extreme tunability of cold-fermion
experiments, which exhibit several experimentally-adjustable
parameters including the interactions, the densities of the
different species, and the trap geometry.  Of late, attention has
focused on one-dimensional (1D) systems with global spin
imbalance~\cite{Yang01, Orso07, HuLiuDrummond, Feiguin, Batrouni,
Zhao,Orso10} or spin-dependent potentials~\cite{Chen10,Zapata},
where the parameter regime occupied by the FFLO state is predicted
to be significantly wider than in the three-dimensional (3D)
case~\cite{SR,Parish07,Yoshida}. Indeed, recent
experiments~\cite{Liao10} on quasi-1D spin-imbalanced fermionic
gases have observed a partially polarized state, although
associated oscillatory pairing correlations have yet to be
confirmed.

In this article, we propose a new 1D setup to achieve FFLO pairing
in cold atomic gases: a {\it balanced\/} mixture of two hyperfine
species of attractively interacting fermionic atoms that are {\it
  separately\/} trapped in a controllable way, as illustrated in
Fig.~\ref{fig:f01}(a)---a situation we call a \lq\lq spin-split
trap.'' This setup provides an effective spatially varying
chemical potential difference between the two spin states due to
the separate trapping potentials and yields an alternate,
dynamically controllable route to achieving oscillatory FFLO-like
pair correlations in cold atomic gases, controlled not by an
imposed global population imbalance but, rather, by the {\it
separation\/} between the two traps and the ensuing local
imbalance.

The spin-split trap, whose 3D counterpart was studied
in Ref.~\cite{Recati06}, is described by the spin-dependent
potentials
\begin{equation}
  V_{\sigma}(z) = \frac{1}{2} m\omega_z^2(z-\sigma d)^2,
  \label{Eq:vupdown}
\end{equation}
where $\omega_z$ is the trapping frequency, $m$ is the atomic mass
and $\sigma=\pm$ corresponds to the two hyperfine species. Thus,
the centers of the two traps are separated by a distance $2d$. For
$d\to 0$, the ground state is a singlet $s$-wave superfluid with a
vanishing spin imbalance everywhere in the cloud.  As argued below
using local density arguments and a Bogoliubov--de Gennes (BdG)
treatment, for nonzero $d$, however, the split traps promote a
local spin imbalance. We find that beyond a critical separation,
$d>d_c$, the split-trap geometry displays oscillatory pairing
correlations, as depicted in Fig.~\ref{fig:f01}(b), which shows
the local pairing amplitude $\Delta(z)$, total density
$\rho(z)=\rho_\uparrow(z)+\rho_\downarrow(z)$, and magnetization
(spin imbalance) $M(z)=\rho_\uparrow(z)-\rho_\downarrow(z)$.

\begin{figure}[t]
  \centering
  \includegraphics[width=7.5cm]{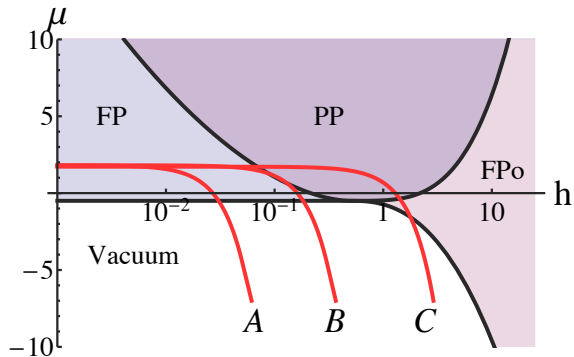}
  \caption{(Color online) The local properties of the system in the spin-split trap
can be understood using the phase diagram of the uniform
imbalanced system, taken from Ref.~\cite{Orso07}, showing fully
paired (FP), partially polarized (PP), and fully polarized (FPo)
phases as well as the vacuum. (Here $\mu$ and $h$ are measured in
units of $mg^2/4 \hbar^2$, where $g$ is the 1D coupling constant.)
The red curves A, B, and C represent the LDA trajectories followed
as a function of $z$ by the spin-split system for $d<d_c$,
$d=d_c$, and $d>d_c$, respectively. } \label{fig:f02}
\end{figure}

\section{Local density approximation}
An intuitive understanding of the spin-split-trap system can be found
using the local density approximation (LDA) along with the known
behavior of the homogeneous spin-imbalanced gas derived using the
Bethe ansatz~\cite{Orso07,HuLiuDrummond}. The phase diagram, shown in Fig.~\ref{fig:f02}, displays three phases as a
function of the net chemical potential
$\mu=(\mu_{\uparrow}+\mu_{\downarrow})/2$ versus the chemical
potential imbalance (magnetic field) $h=(\mu_{\uparrow}-\mu_{\downarrow})/2$, namely a
fully paired (FP) state, a fully polarized (FPo) state, and a
partially polarized (PP) state. The PP state is expected to be of the
FFLO type, having an oscillatory pairing amplitude~\cite{Yang01,Zhao},
as corroborated by our studies below.

Within LDA, the trapping potential in our system enters as a
spin-dependent spatially-varying chemical potential,
$\mu_{\sigma}(z) = \mu_0 - V_\sigma(z)$, where $\mu_0$ is the
global chemical potential of the system. For the harmonic trap of
Eq.~(\ref{Eq:vupdown}), $\mu$ and $h$ are then related through
$\mu=\mu_0-h^2/(2m\omega_z^2 d^2)$, which corresponds to
downward-facing parabolas in the $\mu$ versus $h$ phase diagram.
In Fig.~\ref{fig:f02}, we show three curves corresponding to
different values of the separation $d$. One can see that they
traverse different phases from the center $z=0$ (where $h=0$) to
the edges of the trap. For small separation $d$, the system is
described by a tight parabola and is thus confined to the fully
paired phase, but with increasing $d$, the parabola broadens and,
beyond a critical separation $d_c$, traverses all three phases as
a function of position. In this case, at small $z$, the local
potential imbalance $h$ remains small enough that the system is
(locally) fully paired. At larger $z$, the local $h$ exceeds a
critical value such that (locally) the system enters the PP phase.
At even larger $z$, near the edges of the trap, the system is
(locally) in a fully polarized normal phase. Thus, the system
concurrently hosts all three phases. Note that, in contrast, in
the case of a globally spin-imbalanced system with a single trap,
the system traces a vertical line in the phase diagram, yielding
two regions---a partially polarized core and either fully
polarized or fully paired edges~\cite{Orso07}.

\begin{figure}[t]
  \centering
  \includegraphics[width=8.7cm]{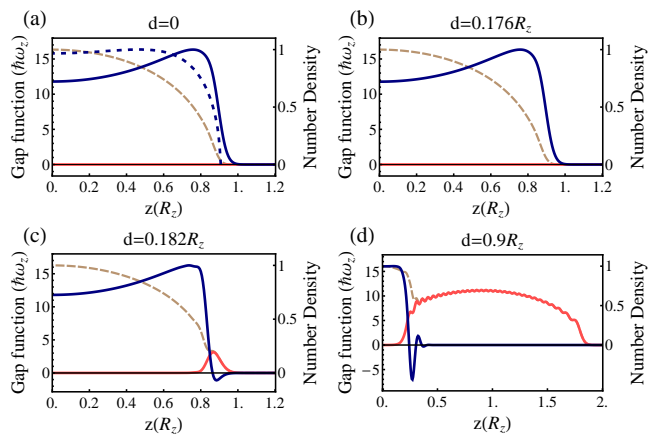}
  \caption{(Color online)(a)--(d) Spatial profile of the gap, total
    density, and magnetization in the $z\ge0$ region [represented as
    in Fig.~\ref{fig:f01}(b)] at $d=0$, $0.176$, $0.182$, and $0.9R_z$,
    respectively. In addition, in (a), the gap
    function obtained by BCS plus LDA is shown (dashed blue curve). In (b) and (c),
    the separations are just below and above the critical value $d_c$ for
    appearance of the first node. Note the scale change on the $z$ axis in (d).}
  \label{fig:f03}
\end{figure}

\section{Microscopic theory}
We now model the spin-split system using a microscopic description
which enables a more detailed analysis, confirms the salient
features described above, and shows a direct correspondence
between local spin imbalance and oscillatory pairing. We study two
species of fermions, $\hat \psi_{\uparrow,\downarrow}(z)$, in a 1D
harmonic potential characterized by the trapping frequency
$\omega_z$. In atomic systems, this limit can be achieved in a
highly anisotropic trap with a transverse trapping frequency
$\omega_r$ such that $N \omega_z / \omega_r < 1$ and
$N|a_{\rm{s}}|/R_z\ll 1$~\cite{Moritz05,Bloch}. Here, $N$ is the
number of fermions of each spin species,
$R_z=\sqrt{2N\!-\!1}\,\ell_z$ (with $\ell_z
=\sqrt{\hbar/m\omega_{z}}$ the oscillator length) is the classical
radius of the free gas in the $z$ direction, and $a_{\rm{s}}$ is
the $s$-wave scattering length for the two-body interactions. The
system is then described by the effective 1D Hamiltonian,
\begin{equation}
  H=\int {dz  \big(\sum\limits_{\sigma} {\hat \psi _\sigma ^\dag
      H_\sigma ^0{{\hat \psi }_\sigma } + g \hat \psi _ \uparrow ^\dag
      \hat \psi _ \downarrow ^\dag {{\hat \psi }_ \downarrow }{{\hat
          \psi }_ \uparrow }} \big) }, \label{eqn:HAM}
\end{equation}
where $H_{\sigma}^0 = -(\hbar^2/2m)
\partial_z^2 + {V_\sigma}(z) - \mu_0$ is the one-particle
Hamiltonian. The 1D coupling constant is given as $g =2 \hbar^2
a_{\rm{s}}/[m \ell_r^2 (1-1.033 a_{\rm{s}}/\ell_r)]$ with the
transverse oscillator length $\ell_r =
\sqrt{\hbar/m\omega_{r}}$~\cite{Olshanii98}.

We analyze our system within the standard BdG treatment, which has
been widely applied to the imbalanced system~\cite{BdG}, taking into
account spin-dependent trapping. The mean-field Hamiltonian, which
self-consistently incorporates the Hartree potential
$U_\sigma=g\langle\hat\psi^\dagger_\sigma\hat\psi_\sigma\rangle$ and
pairing gap
$\Delta=g\langle\hat\psi_\downarrow\hat\psi_\uparrow\rangle$, takes
the form
\begin{equation} {H_M}=\int dz \big[\sum\limits_{\sigma}
  {\hat\psi_\sigma}^\dag(H_\sigma ^0 + {U_\sigma }){\hat \psi_\sigma}
  + (\Delta\hat \psi _ \uparrow ^\dag \hat \psi _ \downarrow ^\dag +
  \rm{H.c.})\big].\label{eqn:MFH}
\end{equation}
We obtain the extended BdG equations in the quasi particle
eigenbasis by a spin-dependent Bogoliubov transformation, $ {{\hat
\psi }_\sigma }({z}) = \sum\limits_n {[{u_{n\sigma }}({z}){{\hat
\gamma }_{n\sigma
    }} - \sigma v_{n\sigma }^*({z})\hat \gamma _{n, - \sigma }^\dag
  ]}$. We use an iterative numerical
procedure~\cite{Richard96} to find self-consistent solutions for
$\rho_\sigma(z)$ and $\Delta(z)$. Parity symmetry between the potentials of the
two species, $V_{\downarrow}(z)=V_{\uparrow}(-z)$, ensures parity
symmetry of the gap function; we find that the even-parity solution, $\Delta(z)=\Delta(-z)$,  is
always energetically favorable. The data presented in the following
were obtained for $N=40$ and $g/\hbar\omega_z R_z=1$.
\begin{figure}[t]
  \centering
  \includegraphics[width=7.5cm]{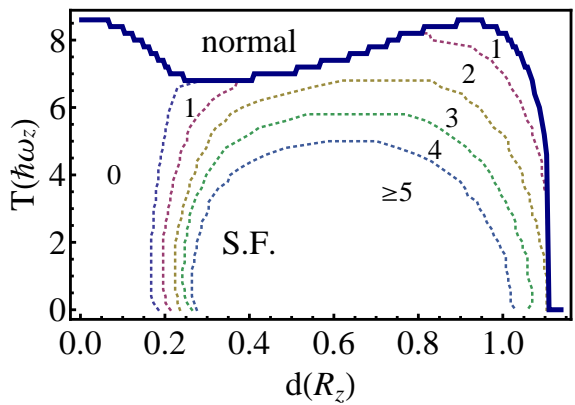}
  \caption{(Color online)Phase diagram as a function of separation and
    temperature ($N=40$, $g/\hbar\omega_z R_z=1$). The solid line separates the normal phase and the
    superfluid phase. In the superfluid phase, the dashed lines
    separate regions of the gap functions with different number of
    nodes.  }\label{fig:f04}\vspace{-.3cm}
\end{figure}

\section{results}
We first focus on the manner in which oscillatory pairing
correlations emerge with increasing separation $d$. In
Fig.~\ref{fig:f03}, we show the pairing gap $\Delta(z)$, total
density $\rho(z)$, and magnetization $M(z)$ for a sequence of four
spin-split-trap systems with increasing $d$.  Panel (a) shows the
$d=0$ case which is fully paired with $M=0$ everywhere, as
expected. The non-monotonicity of $\Delta(z)$ roughly reflects the
functional dependence of the 1D BCS gap on the local chemical
potential $\mu$, that is, $\Delta(z)\propto \mu(z) \exp [-\sqrt
{2{\hbar ^2}{\pi ^2}\mu(z) /(m{g ^2})} ]$. Panel (b) shows that a
small separation, $d<d_c$, does not lead to qualitative changes of
the pairing correlations and the magnetization. Here, the local
$h$ is small enough everywhere that it is energetically favorable
for the system to remain fully paired (i.e., the system is below
the Clogston limit). Panel (c) shows the system just beyond the
critical separation $d_c$, such that, near the edge of the cloud
where the local $h$ is largest and of order $\Delta$, the gap
function $\Delta(z)$ exhibits a node and the magnetization is
finite. As $d$ increases further, the region of oscillatory FFLO
correlations increases and more nodes appear. The progression of
nodes is captured in Figs.~\ref{fig:f01}(b), \ref{fig:f03}(c), and
 \ref{fig:f03}(d). Initially the number of nodes increases as $d$ increases, but
then, beyond a characteristic distance of the order of the cloud
size, diminishes before the system fully separates and becomes
normal.

\begin{figure}[t]
  \centering
  \includegraphics[width=8.5cm]{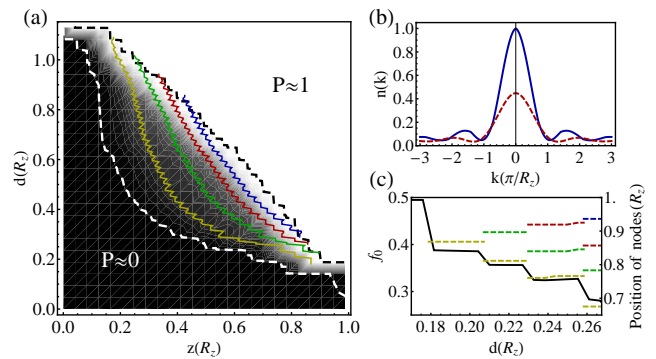}
  \caption{(Color online)(a) Density plot of the polarization $P$ as a function of
    position $z$ and separation $d$.  The gray scale is bounded by 0
    and 1. The dashed white (black) contours correspond to $P=0.01$
    ($0.99$). The solid curves indicate positions of the first four
    nodes. (b) Momentum distribution $n(k)$ for $d=0$ (solid blue
    curve) and $0.25 R_z$ (dashed red curve), normalized by $n(0)$ at
    $d=0$. (c) Fraction of pairs $f_0$ within the central peak [see (b)] of $n(k)$ (solid curve, axis on left-hand side of graph) and
    positions of the first four nodes (dashed curves, axis on
    right-hand side of graph) vs $d$.}
  \label{fig:f05}\vspace{-.3cm}
\end{figure}

We find that the nodal structure is robust against finite
temperature effects.  This is illustrated in the global phase
diagram in Fig.~\ref{fig:f04}, obtained using the parameter values
specified above. Within the superfluid phase, regions with
different numbers of nodes in $\Delta(z)$ are indicated. We note
that the transition temperature in the spatially modulated phase
is of the same order as in the fully paired phase. The number of
nodes decreases with increase in temperature, consistent with the
shrinking of the FFLO region in globally imbalanced
systems~\cite{Wolak}. As for trends with variation of the system
parameters, we numerically find that the critical separation
$\tilde d_c = d_c/R_z$ is independent of $N$ and linearly
dependent on $\tilde g= g/(\hbar \omega_z R_z)$ around $\tilde
g=1$ (in the regime of numerical convergence), which is consistent
with rough estimates based on BCS combined with LDA.

Our results for the behavior of interacting fermions in the
spin-split trap clearly show the intimate connection between a
nonzero polarization and oscillatory pairing correlations. In
Fig.~\ref{fig:f05}(a) we show the polarization,
$P(z)=M(z)/\rho(z)$, as a function of position, $z$, and
separation, $d$, along with the spatial position of the nodes in
$\Delta(z)$. It can be seen that the nodes exist only in the
partially polarized region, $0<P<1$. The correlation between the
polarization and nodal structure indicates that this region is
indeed of the FFLO type, and is surrounded by a fully gapped
superfluid for $P\to 0$ toward the center of the spin-split trap
and a fully polarized normal fluid for $P\to 1$ at the edges.

\section{Experimental aspects}
A direct measure of oscillatory pairing is the pair momentum
distribution function defined as
\begin{equation}
  n(k) =  \int {dz\,dz'e^{ik(z-z')} \big \langle {\hat\psi _
      \uparrow ^\dag (z)\hat\psi _ \downarrow ^\dag (z){\hat\psi _
        \downarrow }(z'){\hat\psi _ \uparrow }(z')} \big \rangle},
  \label{eq:nofk}
\end{equation}
which is experimentally measurable in dynamic-projection
experiments~\cite{PMDF_Homo}. In the homogeneous case, the FFLO
phase is characterized by a peak in $n(k)$ at a characteristic
nonzero wave vector $k$ that depends on the spin
imbalance~\cite{Feiguin,Batrouni}. Typical plots of $n(k)$ in the
spin-split trap are shown in Fig.~\ref{fig:f05}(b) for the cases
of $d=0$ and $d>d_c$.  Due to the spatial inhomogeneity of the
potential imbalance $h$, the system does not possess a
characteristic wavevector. However, $n(k)$ undergoes sudden
changes with increasing separation as Cooper pairs are shifted to
higher momenta.  As shown in Fig.~\ref{fig:f05}(c), the weight
under the central peak suddenly decreases each time a new node
appears in $\Delta(z)$.  Thus, $n(k)$ displays a striking
signature of the modulated phase.

We now turn to the issue of experimentally realizing a
spin-split-trap system. This setup can be achieved via
spin-selective trapping potentials~\cite{Catani,McKay}.
Additionally, a tunable spin-split trap may be achieved using a
magnetic field gradient~\cite{Moritz05,Weld09,Jimenez-Garcia10},
exploiting the distinct hyperfine-Zeeman states of the two fermion
species. To see this, we note that the competition between the
Zeeman effect and hyperfine interaction leads to a nonlinear
energy difference between the two spin states $m_{F\pm}$. We use
the Breit-Rabi formula~\cite{Breit31} to find the
spatially-varying part of the energy difference $\Delta
V(z)=V_\uparrow(z)-V_\downarrow(z)$ in the presence of a field
gradient.  Assuming a spatially-varying field of the form $B(z)
=\bar{B}+B'z$ and expanding the Breit-Rabi formula near the
background field $\bar{B}$, we obtain $\Delta V(z)=2\pi \hbar
B'\tilde{\mu}(\bar B) z$ with $\tilde{\mu}$ the effective \lq\lq
magnetic moment\rq\rq\ given by
\begin{eqnarray}
\tilde\mu(\bar B)=\frac g2\mu_B\sum_{\sigma=\pm}
\sigma\frac{\frac{2m_{F\sigma}}{2I+1}+\frac{\bar B}{B_0}}{\sqrt{1+\frac{4m_{F\sigma}}{2I+1}\frac{\bar B}{B_0}+\frac{\bar B^2}{B_0^2}}}.
\end{eqnarray}
Here, $\mu_B$ is the Bohr magneton, $I$ is the nuclear spin, $B_0$
is the hyperfine field, and $g\simeq 2$.

Using Eq.~(\ref{Eq:vupdown}), we see that a spatial separation $d$
requires a field gradient $B'=2m\omega_z^2d/\tilde{\mu}$. In the
case of interest, we expect that $\bar B$ is close to a Feshbach
resonance (FR) in order to enhance $T_c$ and that $B'$ is small
enough that $a_{\rm{s}}$ can be treated spatially independent in
the system. For $^6$Li ($I=1$, $B_0 = 81$ G), using the hyperfine
levels $m_{F\pm} = -\frac{3}{2}$($+\frac{1}{2}$)~\cite{Schunck}
near the FR at $\bar{B} =691$ G, we find $\tilde\mu_{\rm Li}\simeq
6\times10^{-3}\mu_B$. Assuming a typical trap frequency
$\omega_z\sim 2\pi \times 100$ Hz, a field gradient $B'_{\rm Li}$
of the order of a few hundred G/cm  can achieve a separation $d$
of a few $\ell_z$. (The required field gradient for the more
commonly used two lowest hyperfine levels of $^6$Li is about an
order of magnitude larger and thus much less experimentally
viable.) The most promising case is that of $^{40}$K ($I=4$, $B_0
= -459$ G) with $m_{F\pm} = -\frac{7}{2}$($-\frac{9}{2})$ near the
FR at $\bar{B} = 202$ G. Here $\tilde\mu_{\rm K}\simeq 0.1\mu_B$
and, for the same $\omega_z$ as above, the required gradient
$B'_{\rm K} \simeq 50$ G/cm.

\section{Conclusion}
In summary, we have proposed a novel setting, the spin-split trap,
for observing FFLO-like oscillatory pairing correlations, driven
by a local density imbalance due to the separate trapping
potentials of the two fermion species. Our BdG calculations,
supported by LDA, show that the competition between the tendency
to pair and the tendency towards forming a spin  imbalance leads
to a rich structure that is revealed in quantities such as the
local pairing amplitude and magnetization, as well as in the pair
momentum distribution. Immediate future directions include
investigating the spin-split system through other techniques
amenable to 1D, such as density matrix renormalization group
(DMRG) and quantum Monte Carlo methods, and exploring the exciting
prospect of coupling arrays of spin-split 1D systems.

\section*{ACKNOWLEDGMENTS}
We thank R. Combescot, B. DeMarco, F. Heidrich-Meisner, A.
Lamacraft, A. J. Leggett, C.-H. Pao, R. Hulet, S.-K. Yip, and
M.-H. Yung for helpful discussions. This work was supported by the
NSF under Grants No. DMR-0906521 (S.V.), and No. DMR-0847570
(J.M.) and by the Louisiana Board of Regents, under Grant No.
LEQSF (2008-11)-RD-A-10 (D.S.).  Furthermore, we acknowledge the
hospitality of the Aspen Center for Physics (J.M. and S.V.).

\end{document}